# Panoroma behaviors of Domain walls cosmological models with stability factor in Teleparallel Gravity

A.Y.Shaikh* , A.S.Shaikh ,K.S.Wankhade**

*Department of Mathematics , Indira Gandhi Mahavidyalaya,Ralegaon.India.

*Department of Mathematics ,Poona College of Arts, Science and Commerce,Pune.India.

**Department of Mathematics , Y.C.Science College , Mangrulpir. India.

e-mail:-shaikh_2324ay@yahoo.com

**Abstract:**

The paper is devoted to the study of Locally Rotationally Symmetric (LRS) Bianchi type-I space–time filled with domain walls with volumetric power and exponential expansion laws towards the gravitational field equations for the depiction model $f(T) = T^{\eta}$. The models showing an accelerated expansion with inflationary era in the early and the very late time of the Universe. In power law model, the model is stable at the early phases of the universe and is unstable at late times while in exponential model, the model is stable. The geometrical and physical parameters for both the models are studied. In addition to make the interpretation more clear for that the statefinder diagnostic pair $\{r,s\}$ and jerk parameter are analyzed to characterize different phases of the universe. The well-known astrophysical phenomena, namely the look-back time, proper distance, the luminosity distance and angular diameter distance with red shift are discussed.

1. **Introduction:**

The current accelerating expansion of the universe has experimentally proved by Type Ia Supernovae (Perlmutter et al. (1999), Riess et al.(1999), Efstathiou, et al (2002)) , Cosmic Microwave Background (CMB) radiation (Spergel et al. (2003), Komatsu et al. (2009)), Large Scale Structure (LSS) (Tegmark et al. (2004), Seljak et al. (2005)), Baryon Acoustic Oscillations (BAO) (Eisenstein et al. (2005)) and weak lensing (Jain and Taylor (2003)). Modifying the geometric part of the Einstein-Hilbert action is the best way of exploring dark energy (Magnano et. al.(1987)).Relativists have proposed the probability for the explanation of acceleration of the universe with the help of modified theories (Aktas et. al.(2012)) . The cosmic acceleration of the universe is driven by exotic energy with a large negative pressure which is known as Dark Energy (DE) (Sahni (2004),Frieman et.al.(2008),Anderson et.al.(2013)). Several modified theories of gravitation came into existence such as $f(R), f(R,T), f(G), f(T)$ to explore the dark energy and other cosmological problems. Theory beyond General Relativity (GR) is the Teleparallel Gravity (TG) (Einstein (1928), Einstein (1930)) different from GR (i.e. uses the Weitzenbock connection in place of the Levi-Civita connection) which has no curvature but has torsion and this torsion is responsible for the acceleration of the universe. Ferraro & Fiorini (2007) provided models based on modified TG to inflation. In $f(T)$ gravity, the Teleparallel Lagrangian density described by the function of torsion scalar $T$ in order to account for the late time cosmic acceleration (Bengochea and Ferraro (2009) , Bamba et. al.(2010),Bamba and Geng (2011) and Myrzakulov (2011)). Two new $f(T)$ models were proposed by Linder (2010) in order to explain the present cosmic accelerating expansion. Bamba and Geng (2011) explored the thermodynamics in equilibrium and non-equilibrium descriptions for apparent horizon in $f(T)$ gravity. The validity of Generalized Second Law Thermodynamics on Hubble horizon has been studied by Karami and Abdolmaleki (2012) . Charged wormhole solutions in $f(T)$ gravity with non-commutative background have been extensively explored by Sharif and Rani (2014).The same authors (2015) analyzed the dynamical instability of a spherically symmetric collapsing star in the context of $f(T)$ gravity . $f(T)$ gravity has been extensively studied in the literature by several eminent researchers (Wu and Yu(2010), Yang(2011), Wu and Yu(2011), , Dent, et al.(2011), Li, et al.(2011), Chen, et al.(2011), Bamba, et al.(2011), Wang(2011), Sharif and Rani(2011), Karami and Abdolmaleki(2013) Rodrigues et. al. (2014),Jamil and Yussouf (2015), Abbas et. al.(2015), Gamal and Nashed (2015), Khurshudyan et. al.(2017), Hohmann et. al.(2017), Capozziello et. al. (2017),Channuie and Momeni (2018), Toporensky and Tretyakov(2019).

The cosmological phase transitions in the early universe (Kibble1976) are caused by the topological defects such as cosmic string, domain walls, monopoles and textures associated with spontaneous symmetry breaking. Formations of galaxies are produced due to domain walls and cosmic strings (Vilenkin, 1981;Hill et al., 1989) during a phase transition after the time of recombination of matter and radiation. In the literature survey as specified by Rahaman et. al. (2006), the appearance of domain wall is associated with the breaking of a discrete symmetry i.e. the vacuum manifold $M$ consists of several disconnected components. The space time of cosmological thick domain walls have been studied by Vilenkin (1983), Ipser and Sikivie (1984), Widraw (1989), Goetz (1990),Garfinkle and Gregory (1990), Mukherjee (1993), Wang (1994),Bonjour et al. (1999), Rahaman and Mukherjee (2003) ,Reddy and Rao (2006), Adhav et. al.(2007),Rahaman and Ghosh (2008), Pawar et.al. (2009) ,Katore et.al.(2010a,2010b),Katore et. al.(2011).Sahoo and Mishra (2013) studied quark matter coupled with string cloud and domain walls in the context of general relativity. Biswal et al. (2015) have investigated five dimensional

Kaluza-Klein cosmological models in $f(R,T)$ gravity in presence of domain walls. Kantowski-Sachs space time is considered by Rao and Prasanthi (2016) in $f(R,T)$ gravity in the presence of domain walls. Çatildeglar and Aygün (2016) have obtained exact solutions of strange quark matter attached to string cloud and domain walls for an (n+2) dimensional FRW universe in self-creation cosmology. Bianchi type V cosmological models for domain walls coupled with electromagnetic field in $f(R,T)$ theory have been discussed by Agrawal and Pawar (2017). Lima et. al.(2019) described a scale factor duality (SFD) for domain wall. Pawar et. al. (2020) studied the fractal FRW model within domain wall.

The intellectual urge of curiosity motivates to undertake an investigation into the LRS Bianchi type I cosmological models in the presence of domain walls in the $f(T)$ theory of gravitation.

## 2. Review of $f(T)$ Gravity

The line element for a general space-time metric is defined as

$$dS^2 = g_{\mu\nu}dx^\mu dx^\nu. \tag{1}$$

The tetrads are described as

$$dS^2 = g_{\mu\nu}dx^\mu dx^\nu = \eta_{ij}\theta^i\theta^j, \tag{2}$$

$$dx^\mu = e_i^\mu \theta^i, \quad \theta^i = e_\mu^i dx^\mu, \tag{3}$$

where $\eta_{ij}$ is a metric on Minkowski space-time, $\eta_{ij} = diag[1,-1,-1,-1]$ and

$$e_i^\mu e_\nu^i = \delta_\nu^\mu \text{ or } e_i^\mu e_\mu^j = \delta_i^j.$$

The root of metric determinant is given by $\sqrt{-g} = \det[e_\mu^i] = e$.

The Weitzenböck connection (Aldrovandi and Pereira(2010)) is defined in terms of tetrad field as

$$\Gamma_{\mu\nu}^\alpha = e_i^\alpha \partial_\nu e_\mu^i = -e_\mu^i \partial_\nu e_i^\alpha. \tag{4}$$

The torsion tensor is given by

$$T^{\alpha}_{\mu\nu} = \Gamma^{\alpha}_{\mu\nu} - \Gamma^{\alpha}_{\nu\mu} = e^{\alpha}_i \left(\partial_\mu e^i_\nu - \partial_\mu e^i_\mu\right). \tag{5}$$

The contorsion tensor and superpotential tensor are respectively expressed as

$$K^{\mu\nu}_{\alpha} = \left(-\frac{1}{2}\right)\left(T^{\mu\nu}{}_\alpha - T^{\nu\mu}{}_\alpha - T^{\mu\nu}_\alpha\right) \tag{6}$$

$$S^{\mu\nu}_{\alpha} = \left(\frac{1}{2}\right)\left(K^{\mu\nu}{}_\alpha + \delta^{\mu}_{\alpha} T^{\beta\nu}{}_\beta - \delta^{\nu}_{\alpha} T^{\beta\mu}_\beta\right). \tag{7}$$

The torsion scalar $T$ is specified by

$$T = T^{\alpha}_{\mu\nu} S^{\mu\nu}_{\alpha}, \tag{8}$$

The action of teleparallel gravity is defined as

$$S = \int [T + f(T) + L_{matter}] e\, d^4x. \tag{9}$$

Using Eq.(9), one can obtain the following field equation

$$S^{\nu\rho}_{\mu}\partial_{\rho}T f_{TT} + \left[e^{-1} e^{i}_{\mu}\partial_{\rho}\left(e e^{\alpha}_i S^{\nu\rho}_{\alpha}\right) + T^{\alpha}{}_{\lambda\mu} S^{\nu\lambda}_{\alpha}\right](1 + f_T) + \frac{1}{4}\delta^{\nu}_{\mu}(T + f) = k^2 T^{\nu}_{\mu}, \tag{10}$$

where $T^{\nu}_{\mu}$ is the energy momentum tensor of the matter source, while $f_T$ and $f_{TT}$ denote the first and second derivatives of the function $f(T)$ with respect to $T$.

## 3. Metric Space and field equations

Anisotropy plays a significant role in the early stage of evolution of the universe which are very well explained by Bianchi type models. In this work, we are interested to explore $f(T)$ gravity using Locally Rotationally Symmetric (LRS) Bianchi type I space-time. The line element is of the form

$$ds^2 = dt^2 - A^2(t)dx^2 - B^2(t)\left[dy^2 + dz^2\right], \tag{11}$$

where $A$ and $B$ be the functions of cosmic time $t$ only.

The corresponding Torsion scalar is given by

$$T = -2\left(2\frac{\dot{A}\dot{B}}{AB} + \frac{\dot{B}^2}{B^2}\right). \tag{12}$$

The energy momentum tensor of the domain walls in conventional form (Banerji and Das (1988)) is given by

$$T_{ij} = \rho(g_{ij} + \omega_i\omega_j) + pg_{ij}, \tag{13}$$

where $\rho$ is the energy density of the wall, $p$ is the pressure in the direction normal to the plane of the wall and $\omega_i$ is a unit space-like vector in the same direction (Rahaman et al. 2001) . The energy momentum tensor of domain walls contains the normal matter $\rho_m, p_m$ and the tension $\eta$ ( Vilenkin (1981)). The relation is given by

$$p = p_m - \eta, \rho = \rho_m + \eta \text{ where } p_m = (\gamma - 1)\rho_m, 1 \leq \gamma \leq 2. \tag{14}$$

In co-moving coordinate system $\omega^i = (0,0,0,1)$ satisfying $\omega^i\omega_i = -1$. Using equation (13), the non-vanishing components of $T_{ij}$ can be obtained as

$$T_1^1 = T_2^2 = T_3^3 = \rho, T_4^4 = -p \quad , T_j^i = 0, i \neq j, \tag{15}$$

where $p$ and $\rho$ are the functions of time 't' only.

Using equations (10), (11) ,(13) and (15) , one can obtain the field equations as

$$(T+f) + 4(1+f_T)\left\{\frac{\ddot{B}}{B} + \frac{\dot{B}^2}{B^2} + \frac{\dot{A}\dot{B}}{AB}\right\} + 4\frac{\dot{B}}{B}\dot{T}f_{TT} = k^2\rho, \tag{16}$$

$$(T+f) + 2(1+f_T)\left\{\frac{\ddot{A}}{A} + \frac{\ddot{B}}{B} + \frac{\dot{B}^2}{B^2} + 3\frac{\dot{A}\dot{B}}{AB}\right\} + 2\left\{\frac{\dot{A}}{A} + \frac{\dot{B}}{B}\right\}\dot{T}f_{TT} = k^2\rho, \tag{17}$$

$$(T+f) + 4(1+f_T)\left\{\frac{\dot{B}^2}{B^2} + 2\frac{\dot{A}\dot{B}}{AB}\right\} = -k^2 p. \tag{18}$$

where the dot($\cdot$) denotes the derivative with respect to time $t$.

The set of field equations (16)-(18) is highly non-linear differential equations containing five unknowns namely $A, B, f, p, \rho$ with three field equations.

The spatial volume is given by

$$V = R^3 = AB^2, \qquad (19)$$

where the average scale factor $R$ can be written in terms of metric functions.

The mean Hubble parameter for LRS B-I can be expressed as

$$H = \frac{1}{3}(H_1 + H_2 + H_3), \qquad (20)$$

where $H_1 = \frac{\dot{A}}{A}$, $H_2 = H_3 = \frac{\dot{B}}{B}$ are the directional Hubble parameter in the direction of $x$, $y$ and $z$-axis respectively.

The relation between the parameters $H, V$ and $R$ is given by :

$$H = \frac{1}{3}\frac{\dot{V}}{V} = \frac{1}{3}(H_1 + H_2 + H_3) = \frac{\dot{R}}{R}. \qquad (21)$$

Expansion scalar and shear scalar are expressed as

$$\theta = u^{\mu}_{;\mu} = \frac{\dot{A}}{A} + 2\frac{\dot{B}}{B}, \qquad (22)$$

$$\sigma^2 = \frac{3}{2}H^2 A_m. \qquad (23)$$

The mean anisotropic parameter $A_m$ is defined as

$$A_m = \frac{1}{3}\sum_{i=1}^{3}\left(\frac{H_i - H}{H}\right)^2. \qquad (24)$$

The deceleration parameter is defined and expressed as

$$q = \frac{d}{dt}\left(\frac{1}{H}\right) - 1. \qquad (25)$$

## 4. Solution of the Field Equations

To obtain exact solution of the field equations, the depiction model of $f(T)$ is considered i.e.

$$f(T) = T^\eta. \qquad (26)$$

Using equations (16) and (17), it yields

$$\frac{d}{dt}\left(\frac{\dot{A}}{A} - \frac{\dot{B}}{B}\right) + \left(\frac{\dot{A}}{A} - \frac{\dot{B}}{B}\right)\frac{\dot{V}}{V} = 0 \qquad (27)$$

which on integration gives

$$\frac{A}{B} = k_2 \exp\left[k_1 \int \frac{dt}{V}\right] \qquad (28)$$

where $k_1$ and $k_2$ are constants of integration.

In view of equation (19), one can write $A$ and $B$ in the explicit form

$$A = D_1 V^{\frac{1}{3}} \exp\left(\chi_1 \int \frac{1}{V} dt\right), \qquad (29)$$

$$B = D_2 V^{\frac{1}{3}} \exp\left(\chi_2 \int \frac{1}{V} dt\right), \qquad (30)$$

where $D_i \, (i=1,2)$ and $\chi_i \, (i=1,2)$ satisfying the relation $D_1 D_2^2 = 1$ and $\chi_1 + 2\chi_2 = 0$. Since the field equations are highly nonlinear, an extra condition is needed to solve the system completely. In this paper, the two different volumetric expansion laws are considered found in the literature.

$$V = at^n \qquad (31)$$

and

$$V = \alpha e^{mt}, \qquad (32)$$

where *a, n,* $\alpha, m$ are constants. In this way, all possible expansion histories, the power-law expansion (31) and the exponential expansion (32) have been covered.

## 5. Model for power law

Using equations (29),(30) and (31), the metric potentials are expressed as

$$A = D_1 a^{\frac{1}{3}} t^{\frac{n}{3}} \exp\left\{\frac{\chi_1}{a(1-n)} t^{1-n}\right\} \tag{33}$$

and

$$B = D_2 a^{\frac{1}{3}} t^{\frac{n}{3}} \exp\left\{\frac{\chi_2}{a(1-n)} t^{1-n}\right\}. \tag{34}$$

The metric potentials *A* and *B* vanish at an initial epoch possessing initial singularity and at large times i.e. $t \to \infty$, both the scale factors *A* and *B* tend to infinity which is in complete agreement with the Big-Bang model of the universe (Shaikh and Katore (2015)).

Thus the metric (11) filled with matter and holographic dark energy fluid in teleparallel gravity becomes

$$ds^2 = dt^2 - D_1^2 a^{\frac{2}{3}} t^{\frac{2n}{3}} \exp\left\{\frac{2\chi_1}{a(1-n)} t^{1-n}\right\} dx^2 - D_2^2 a^{\frac{2}{3}} t^{\frac{2n}{3}} \exp\left\{\frac{2\chi_2}{a(1-n)} t^{1-n}\right\} (dy^2 + dz^2). \tag{35}$$

The Torsion scalar for the model becomes

$$T = \frac{-2n^2}{3t^2} - \frac{2\chi_2(2\chi_1 + \chi_2)}{a^2 t^{2n}}. \tag{36}$$

The mean Hubble's parameter is given by

$$H = \frac{n}{3t}. \tag{37}$$

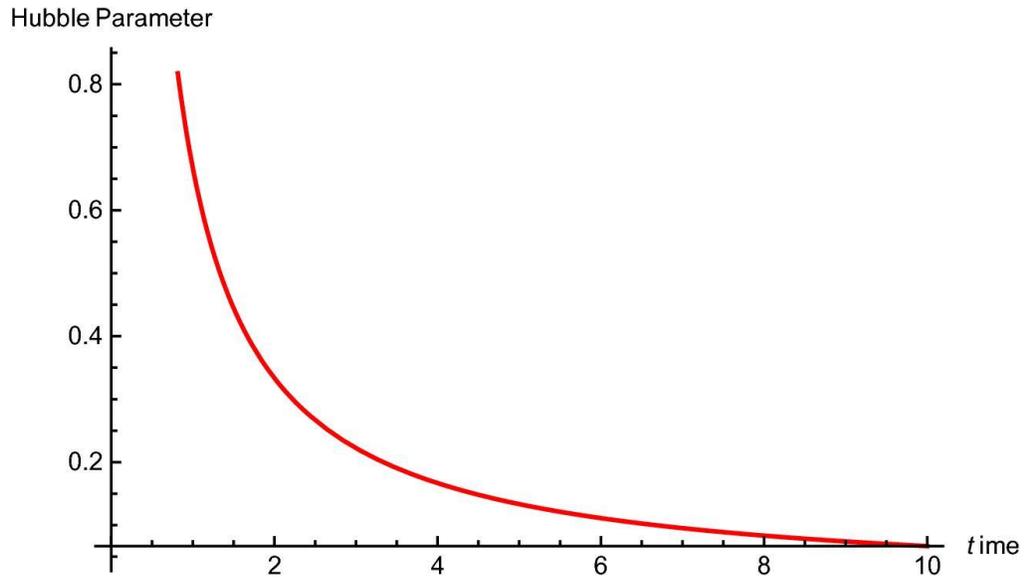

Figure 1. Hubble Parameter vs time for $n = 1.5$.

Expansion scalar is given by

$$\theta = \frac{n}{t}. \tag{38}$$

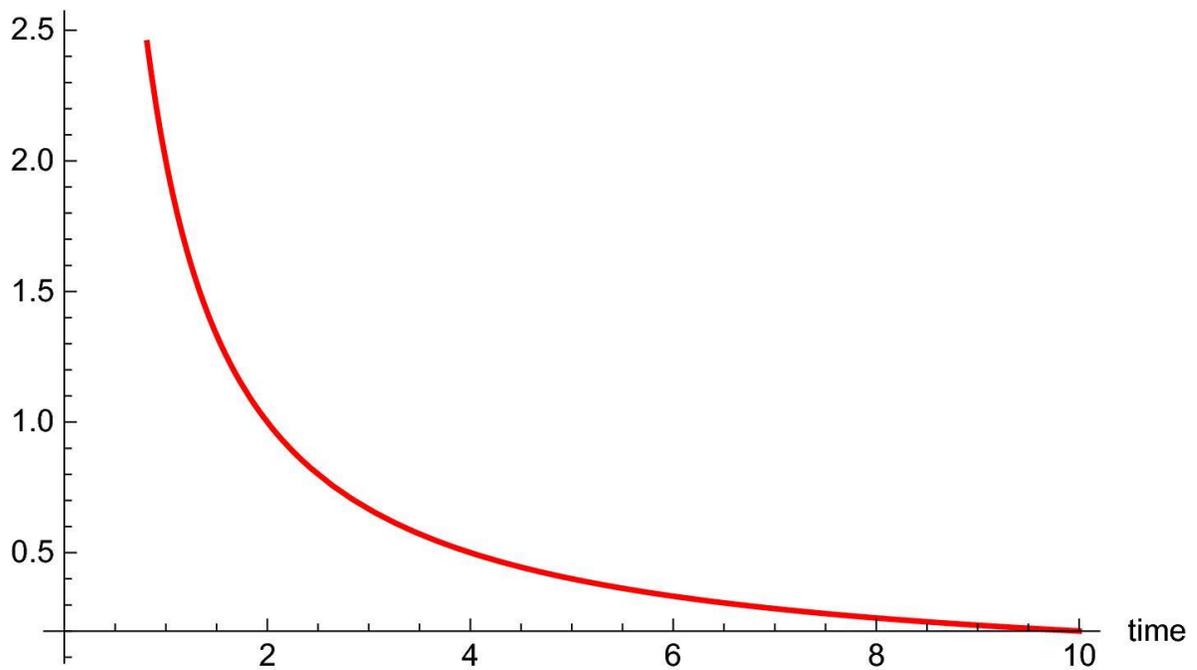

Figure 2. Scalar Expansion vs time for $n = 1.5$.

Shear scalar is

$$\sigma^2 = \frac{\chi^2}{2a^2 t^{2n}}. \tag{39}$$

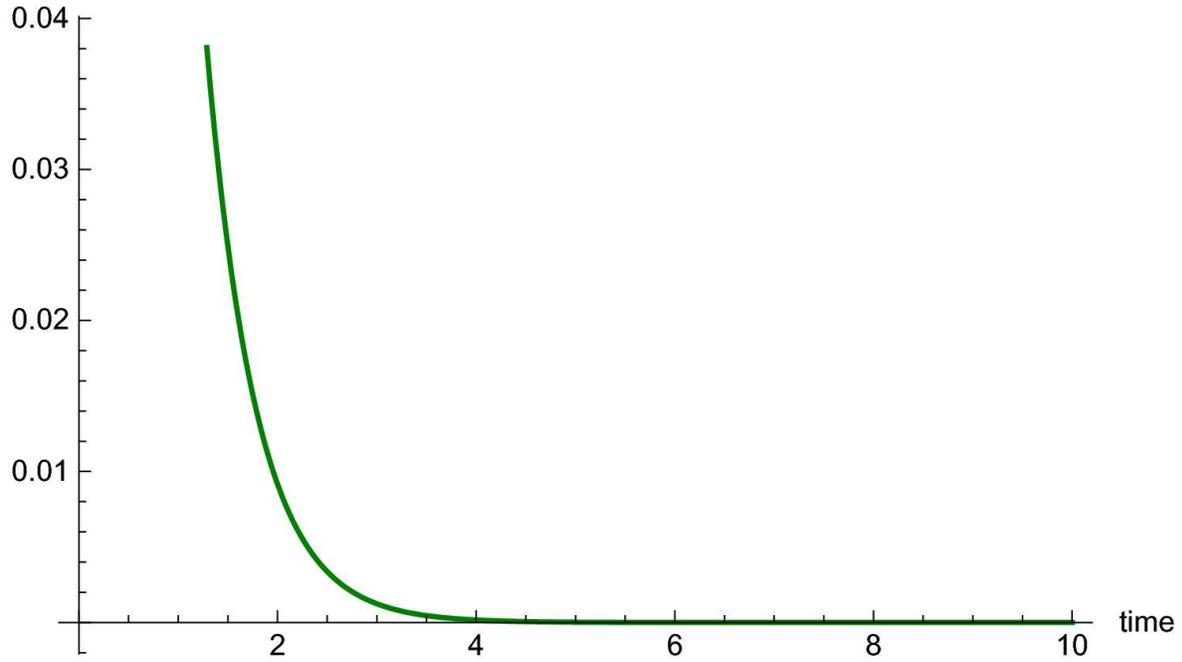

Figure 3. Shear Scalar vs time for $\alpha = 1, \chi = 0.05, n = 1.5$.

The mean anisotropy parameter is

$$\Delta = \frac{3\chi^2}{a^2 n^2 t^{2(n-1)}}, \tag{40}$$

where $\chi^2 = 2\chi_1^2 + \chi_2^2 =$ constant .

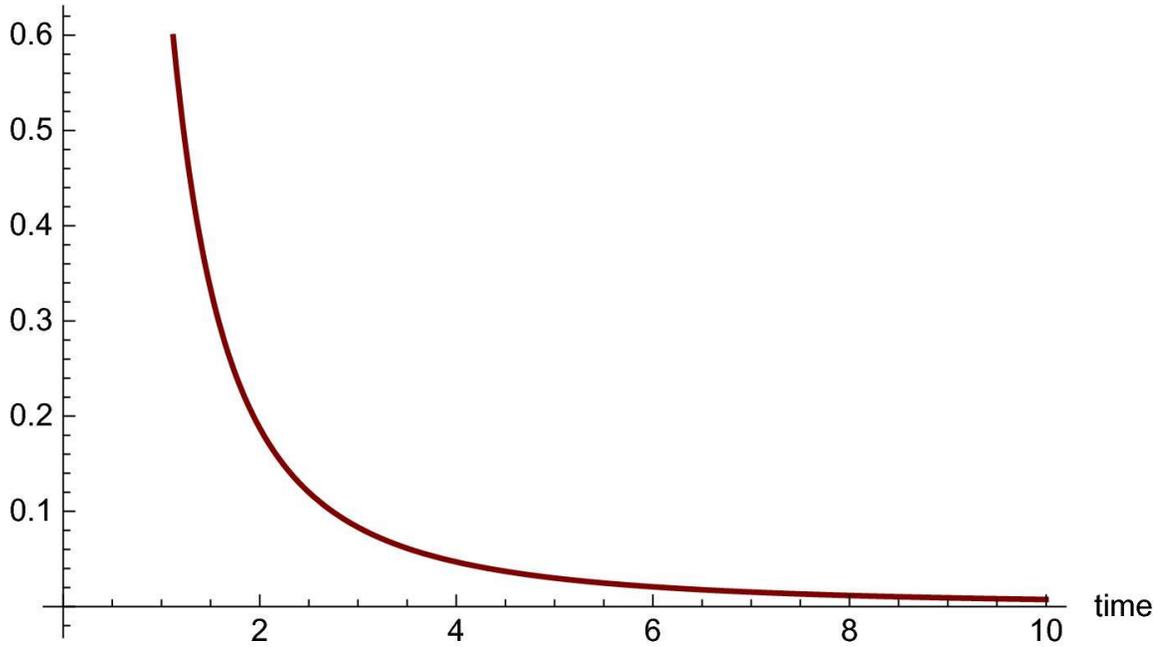

Figure 4. Anisotropic Parameter vs time for $\alpha=1, \chi=0.05, n=1.5$.

The deceleration parameter is obtained as

$$q = \frac{3}{n} - 1. \tag{41}$$

Spatial volume of the universe starts with big bang at an initial epoch, expands with the increase of time and hence there is a possibility of inflation. This shows that the universe starts evolving with zero volume and expands with time *t*. It is observed that the Torsion of the Universe is time dependent. The mean Hubble parameter is a decreasing function of time. It is very large at the initial epoch and tends to zero at late times (figure 1). From figure 2, it is observed that $\theta \to \infty$ for $t \to 0$ and tend to zero as time becomes infinitely large. The shear scalar diverge at $t=0$. As $t \to \infty$, shear scalar tend to zero as shown in figure 3. Therefore, the models would essentially give an empty universe for large times *t* i.e. at late time matter has no shear. From figure 4, it is observed that mean anisotropic parameter is very large at $t=0$ and tends to zero as $t \to \infty$. This faster decrement may be due to the presence of domain walls which means that initially the universe is anisotropic and it tends to isotropy at late times. Therefore, the solutions reveal that domain walls have played an important role in the process of isotropization of the large scale structure of the universe. The model inflates or not depends on the indication of the sign of $q$. A positive sign of $q$ corresponds to the standard decelerating model whereas the negative sign of $q$ indicates inflation. The deceleration parameter is negative for $n>3$ i.e. the universe is accelerating (Ade et. al. (2013)) indicating that the universe is fast and the value of the deceleration parameter lies somewhere in the range $-1<q<0$ which is in agreement with current observations of SNe Ia and CMB.

The pressure of the Domain Walls

$$p = -\frac{1}{k^2}\left\{\frac{4n^2}{3t^2} + \frac{4\chi_2(2\chi_1 + \chi_2)}{\alpha_1 t^{2n}}\right\}. \tag{42}$$

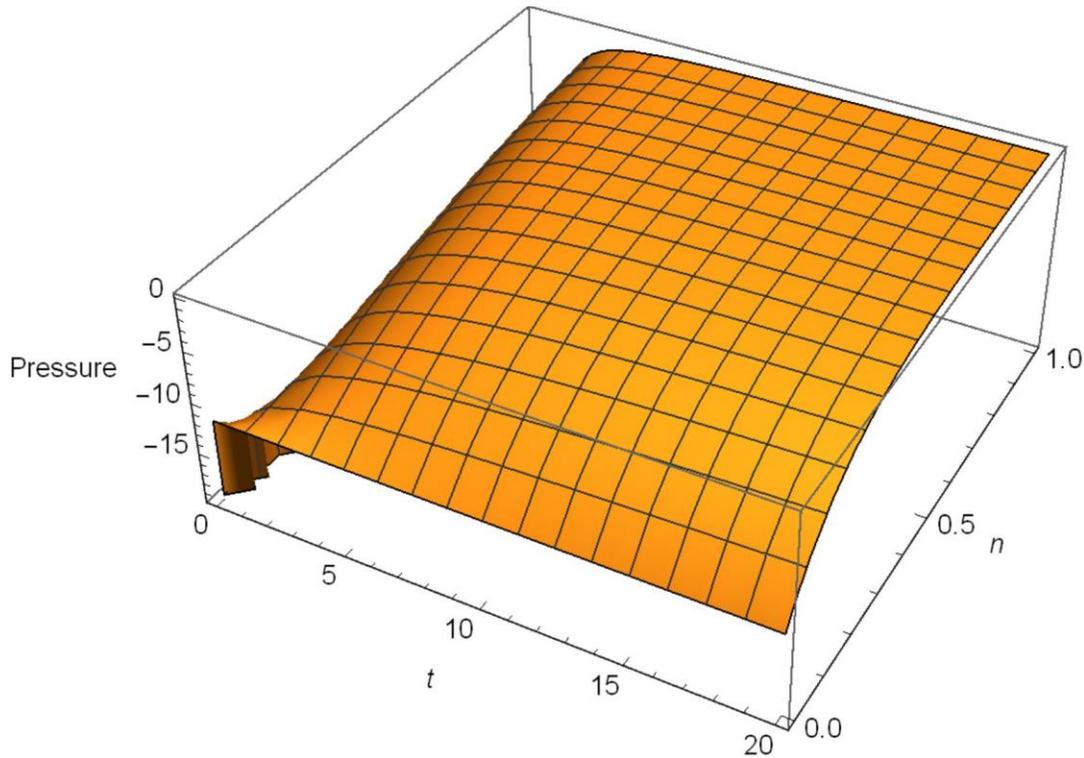

Figure 5. Pressure vs time and $n$ for $\alpha_1 = 1, \chi_1 = \chi_2 = 0.05, k = 0.5$.

Energy Density of Domain Walls

$$\rho = \frac{1}{k^2}\left\{\frac{12n(n-2)}{9t^2} + \frac{12\chi_2^2}{\alpha_1^2 t^{2n}}\right\}. \tag{43}$$

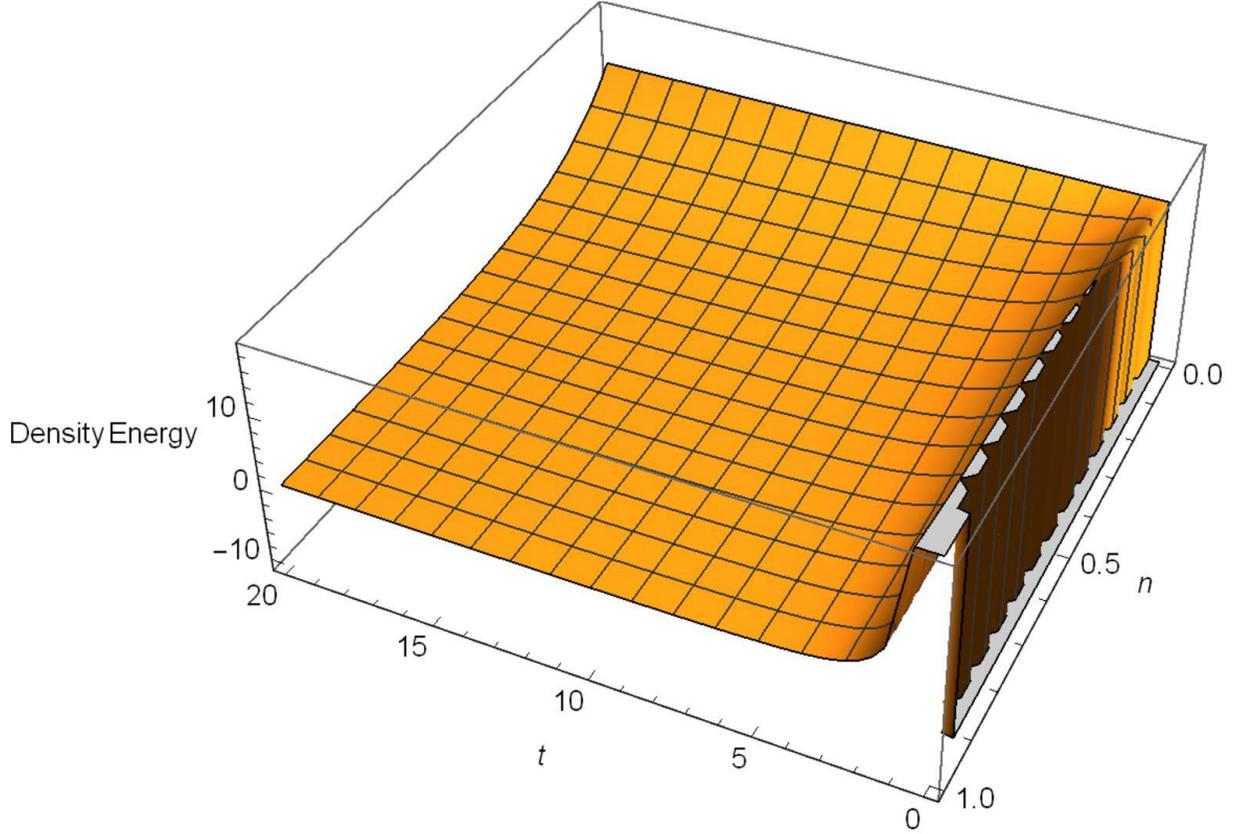

Figure 6. Energy density of domain walls vs time and $n$ for $\alpha_1 = 1, \chi_1 = \chi_2 = 0.05, k = 0.5$.

Energy density of the matter

$$\rho_m = \frac{1}{\gamma k^2}\left\{\frac{12n(n-2)}{9t^2} + \frac{12\chi_2^2}{\alpha_1^2 t^{2n}} - \frac{4n^2}{3t^2} - \frac{4\chi_2(2\chi_1 + \chi_2)}{\alpha_1 t^{2n}}\right\}. \tag{44}$$

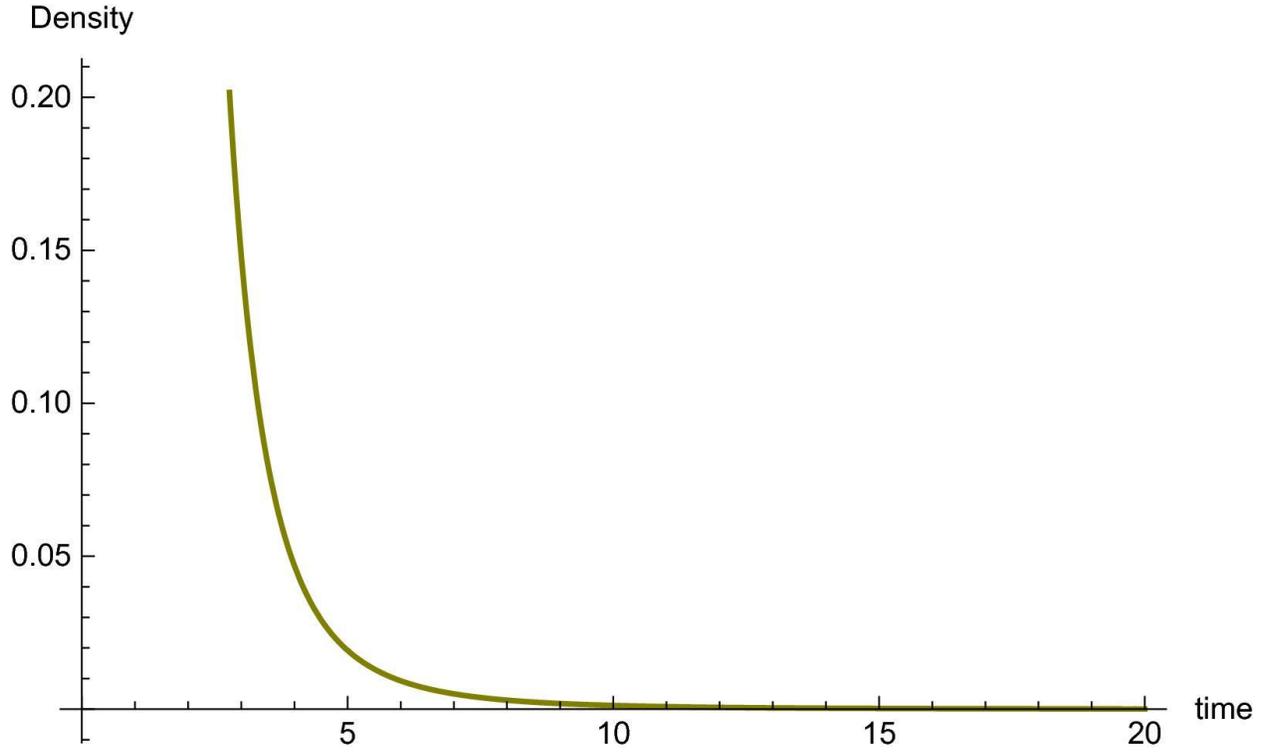

Figure 7. Energy density of matter vs time for $\alpha_1 = 1, \chi_1 = \chi_2 = 0.05, k = 0.5, n = 1.5, \gamma = 0.33$.

Tension $\eta$ of the Domain wall as

$$\eta = \frac{1}{\gamma k^2}\left\{(\gamma-1)\left[\frac{12n(n-2)}{9t^2} + \frac{12\chi_2^2}{\alpha_1^2 t^{2n}}\right] + \frac{4n^2}{3t^2} + \frac{4\chi_2(2\chi_1 + \chi_2)}{\alpha_1 t^{2n}}\right\} \qquad (45)$$

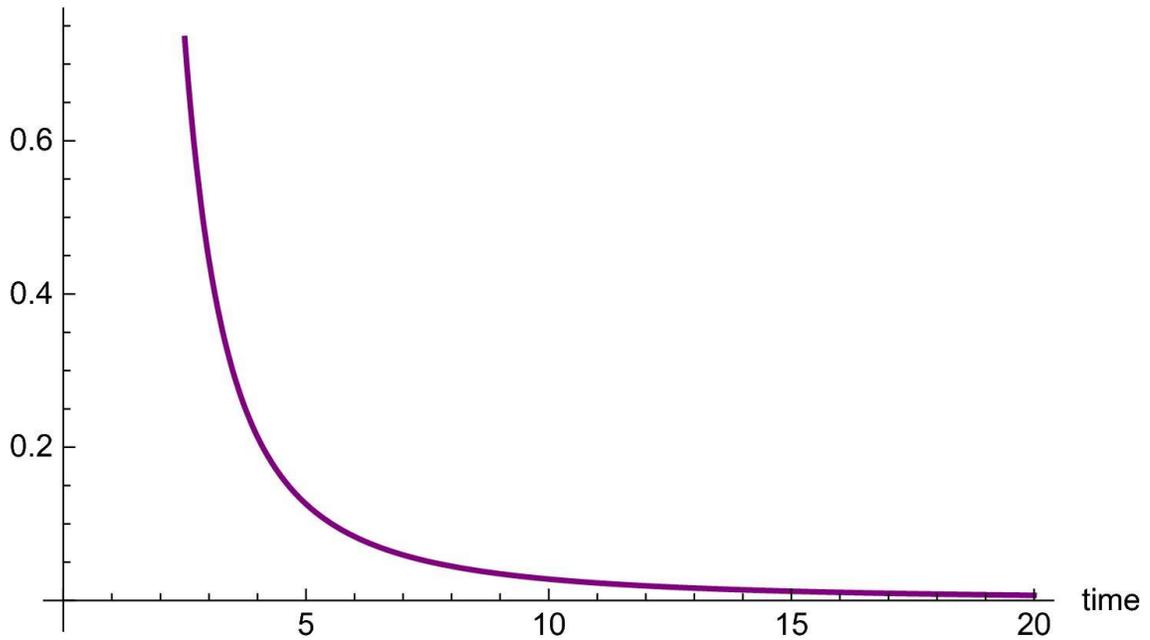

Figure 8. Tension of Domain walls vs time for $\alpha_1 = 1, \chi_1 = \chi_2 = 0.05, k = 0.5, n = 1.5, \gamma = 0.33$.

It is clear from figure 5 that pressure assumes negative values throughout the evolution of the cosmic time. A negative pressure is required to provide an antigravity effect and to drive the acceleration .Thus our derived model is evidence from the host of observational data favoring an accelerated expansion of the universe. It is observed from figure 6 that $\rho \to \infty$ as $t \to \infty$ i.e. there is gravitational collapse at late times and it occurs due to the action of the tension of the curved walls (Zel'dovich et. al.(1975)).

The energy density of the matter has been graphed versus time in figure 7. It is observed that the energy density of matter remains always positive and decreasing function of time i.e. tends to zero as time increases indefinitely and also possess initial singularity at an initial epoch. Thus the universe approaches towards a flat universe at late time which is in good agreement with the recent observational data.

The tension of the domain wall is decreasing function of cosmic time. It is positive and maximum near the Big Bang i.e. $t = 0$, hence the domain walls could exist at an early era of the evolution of the universe. The tension of the domain walls vanishes at infinite time, as shown in figure 8, which is in accordance with Zeldovich et al.(1975). Thus the existence of domain walls at later time leads to the inhomogeneous universe, which contradicts the present observational data.

## 6. Model for exponential law

Using equations (29),(30) and (32), the metric potentials are given as :

$$A = D_1 \alpha^{\frac{1}{3}} e^{\frac{mt}{3}} \exp\left\{\frac{-\chi_1}{\alpha m} e^{-mt}\right\} \tag{46}$$

and

$$B = D_2 \alpha^{\frac{1}{3}} e^{\frac{mt}{3}} \exp\left\{\frac{-\chi_2}{\alpha m} e^{-mt}\right\}. \tag{47}$$

Thus the metric (11) filled with matter and holographic dark energy fluid within the framework of $f(T)$ gravity becomes

$$ds^2 = dt^2 - D_1^2 \alpha^{\frac{2}{3}} e^{\frac{mt}{3}} \exp\left\{-2\left[\frac{\chi_1}{\alpha m} e^{-mt}\right]\right\}(dx^2) - D_2^2 \alpha^{\frac{2}{3}} e^{\frac{mt}{3}} \exp\left\{-2\left[\frac{\chi_2}{\alpha m} e^{-mt}\right]\right\}(dy^2 + dz^2). \tag{48}$$

The Torsion scalar for the model becomes

$$T = -2\left\{\frac{m^2}{3} + \frac{\chi_2(2\chi_1 + \chi_2)e^{-2mt}}{\alpha^2}\right\}. \tag{49}$$

The mean Hubble's parameter is given by

$$H = \frac{m}{3}. \tag{50}$$

The anisotropy parameter of the expansion is

$$\Delta = \frac{3\chi^2 e^{-2mt}}{\alpha^2 m^2}. \tag{51}$$

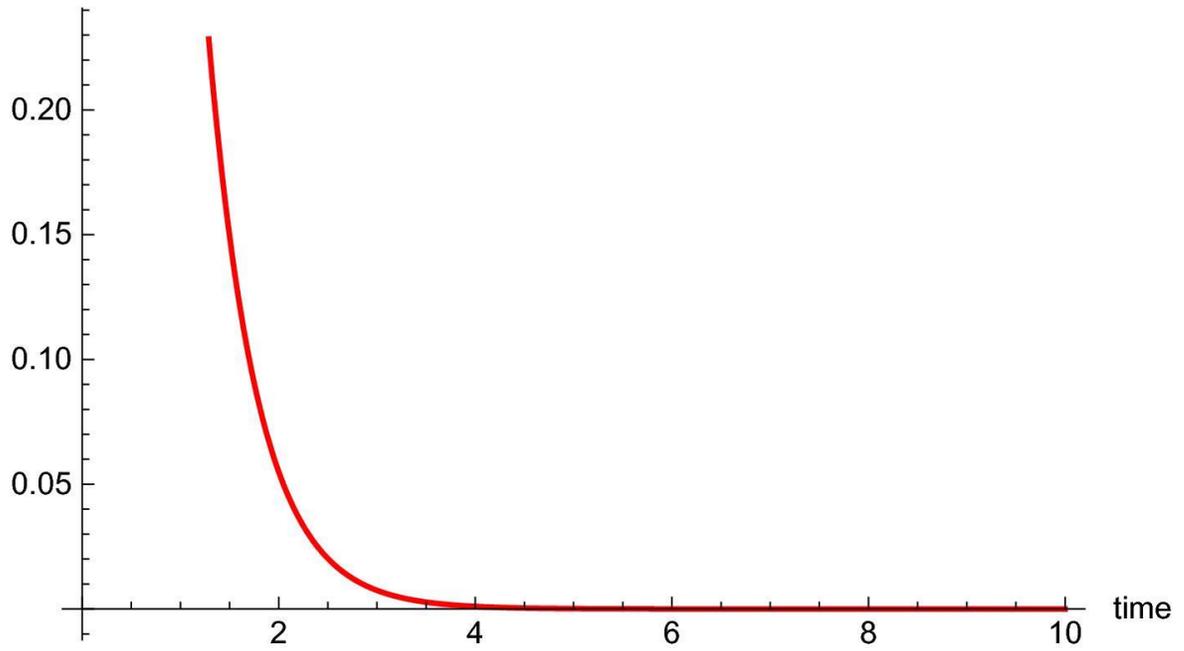

Figure 9. Anisotropy parameter vs time for $\alpha = 1, \chi = 0.05, m = 2.0$.

The expansion scalar yields

$$\theta = m. \tag{52}$$

The shear scalar is obtained as

$$\sigma^2 = \frac{\chi^2 e^{-2mt}}{2\alpha^2}. \tag{53}$$

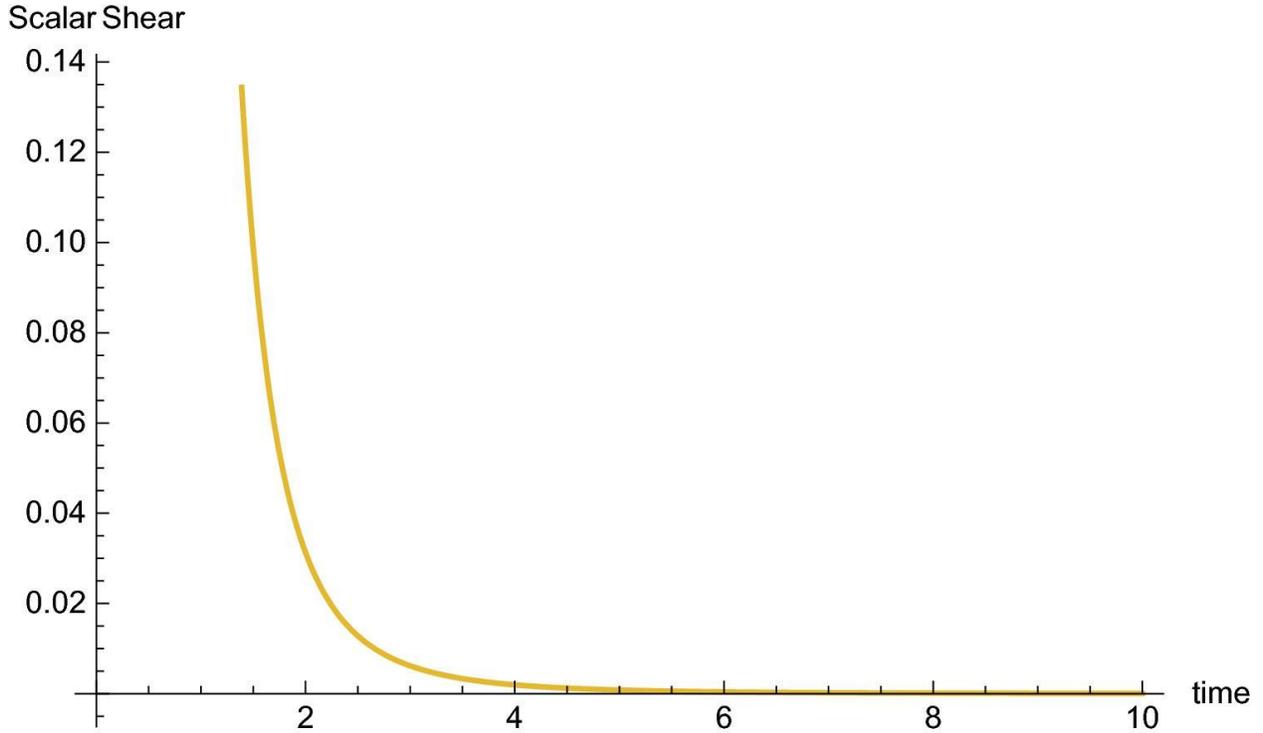

Figure 10. Shear Scalar vs time for $\alpha = 1, \chi = 0.05, m = 2.0$.

The deceleration parameter

$$q = -1. \tag{54}$$

The scale factors admit constant values at an initial epoch imply that the model has no singularity and at far future, they diverge to infinity. Thus the volume of the universe is an exponential function which expands with increase in time from a constant to infinitely large. This is consistent with big bang scenario. The anisotropy parameter is constant at an initial epoch i.e the universe was anisotropic at early stage and expands isotropically at later times as depicted in figure 9. The expansion scalar $\theta$ remains constant throughout the evolution. The rate of expansion of the universe is constant for $m > 0$. Thus, the universe evolves with constant rate of expansion. The ratio of the shear scalar to the expansion scalar was non-zero and as the time increases, it tends to be zero at the early stages of the evolution of the universe, which means that the universe was initially anisotropic and at a late time it approaches isotropy. From figure 10 , it can be seen that the shear scalar $\sigma \to 0$ as $t \to \infty$. The obtained model is consistent with the present scenario of accelerating expansion of the universe (Riess et al. 1998; Perlmutter et al. 1999), as the deceleration parameter $q$ has negative value . The spatial volume $V$ is finite at $t=0$, expands exponentially as $t$ increases and becomes infinitely large as $t \to \infty$. Here we have $\frac{dH}{dt}=0 \Rightarrow q=-1$ which provides the quickest rate of growth of the universe and may represent the inflationary era in the early universe and the very late time of the universe.

Pressure of the Domain Walls

$$p = -\frac{1}{k^2}\left\{\frac{4m^2}{3} + \frac{4\chi_2(2\chi_1 + \chi_2)e^{-2mt}}{\alpha_2^2}\right\}. \tag{55}$$

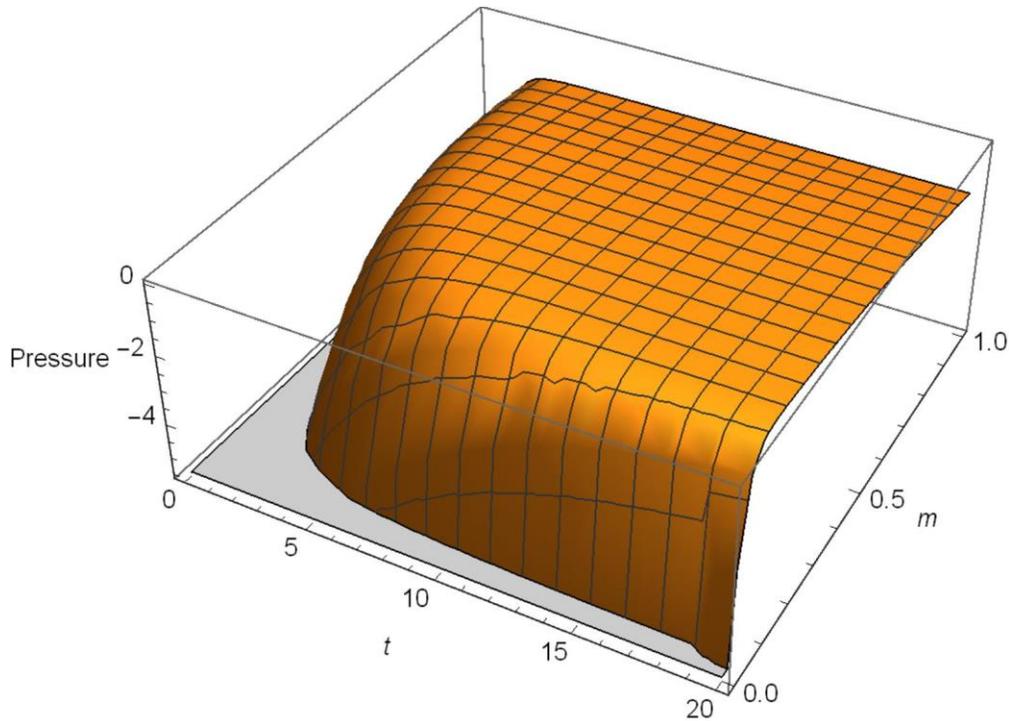

Figure 11. Pressure vs time and $m$ for $\alpha_2 = 1, \chi_1 = \chi_2 = 0.05, k = 0.5$.

Energy Density of Domain Walls

$$\rho = \frac{1}{k^2}\left\{\frac{4m^2}{3} - \frac{4\chi_2(2\chi_1 + \chi_2)e^{-2mt}}{\alpha_2^2}\right\}. \tag{56}$$

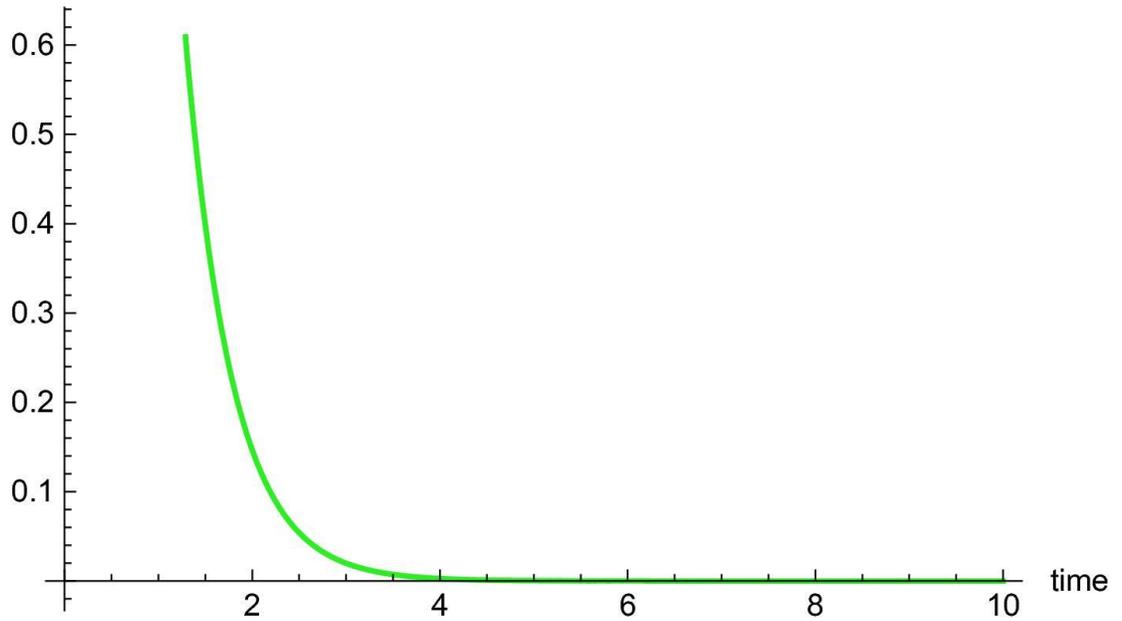

Figure 12. Energy Density vs time for $\alpha_2 = 1, \chi_1 = \chi_2 = 0.05, m = 2.0, k = 0.5$.

The energy density of normal matter is obtained as

$$\rho_m = \frac{8}{\gamma k^2}\left\{\frac{\chi_2(2\chi_1 + \chi_2)e^{-2mt}}{\alpha_2^2}\right\} \tag{57}$$

The tension of the Domain Walls as

$$\eta = \frac{4}{k^2}\left\{\frac{m^2}{3} + \frac{(2-\gamma)}{\gamma}\frac{\chi_2(2\chi_1 + \chi_2)e^{-2mt}}{\alpha_2^2}\right\} \tag{58}$$

Domain Tension walls

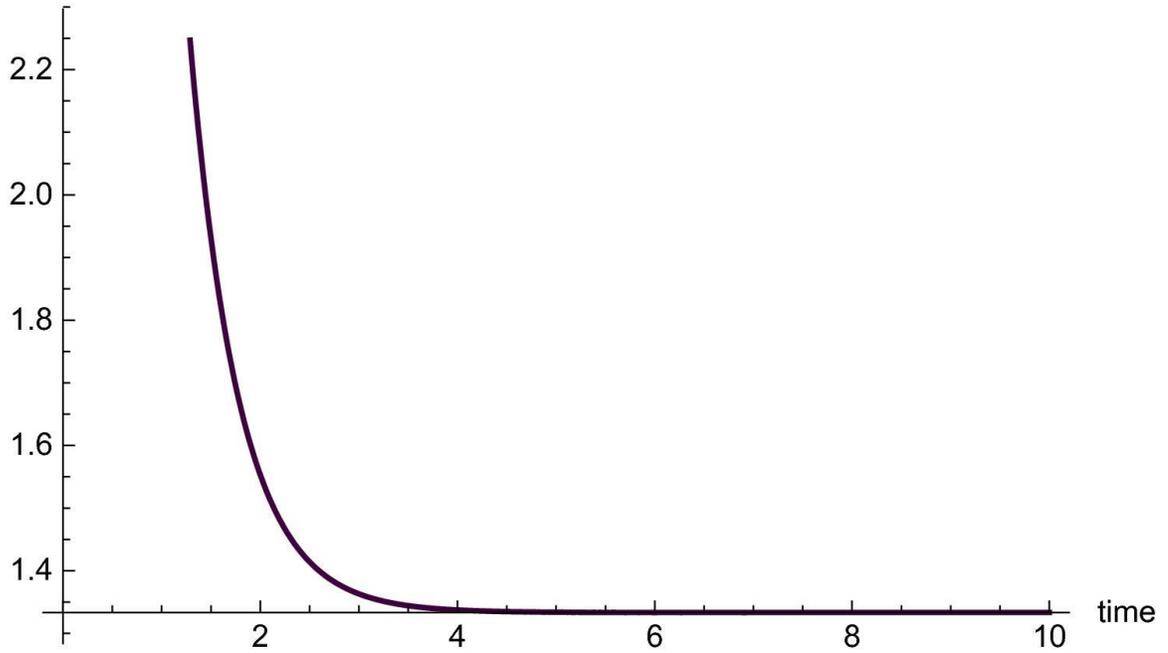

Figure 13. Tension density of domain walls vs time for $\alpha_2 = 1, \chi_1 = \chi_2 = 0.05, m = 2.0, k = 0.5$.

The pressure is negative throughout the evolution as shown in figure 11. The existence of dark energy in the universe is due to the fact of strong negative pressure (Shamir (2014)). From Fig 12, it can be seen that the energy density of domain walls ($\rho$) is decreasing function of time. It is large at an initial epoch i.e near Big-Bang singularity but starts decreasing monotonically with the passage of time and tends to be very small at large times. It is observed that the energy density is the decreasing function of time. It is very large at $t = 0$ and tends to zero for large value of $t$. From figure 13, it is observed that the tension of the domain walls is decreasing function of cosmic time. It is positive and extreme large near the Big Bang singularity and tends to be zero at infinite value of cosmic time which is consistent with the investigations of Zel'dovich et al. (1975). The tension of the domain walls is totally different with the results obtained by Yilmaz (2006) and Katore et al (2010) in different context.

## 7. Stability of the model

For the stability of the models, the function $v_s = \dfrac{dp}{d\rho}$ is utilized to check whether our models are physically acceptable. The model is stable when $\dfrac{dp}{d\rho} > 0$ (Sadeghi et.al.(2013)) and for $\dfrac{dp}{d\rho} < 0$, it is an unstable model.

### 7.1. The power law model

$$v_s = \frac{dp}{d\rho} = -\frac{\left(\dfrac{8n^2}{3t^3} + \dfrac{8n\chi_2(2\chi_1 + \chi_2)}{\alpha_1 t^{2n+1}}\right)}{\left(\dfrac{24n(n-2)}{9t^3} + \dfrac{24n\chi_2^2}{\alpha_1^2 t^{2n+1}}\right)} \tag{59}$$

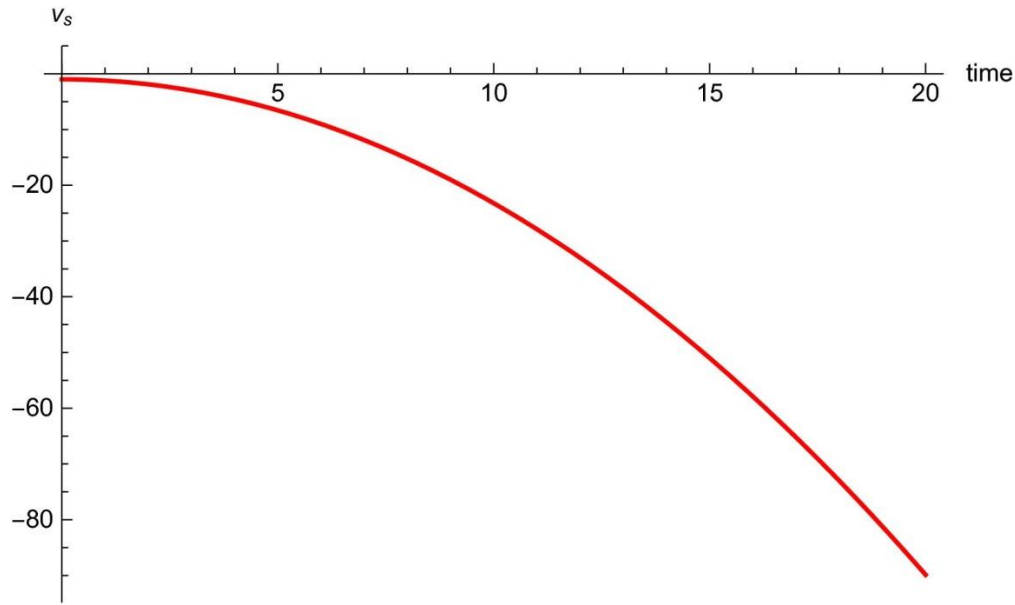

Figure 14 . $v_s$ vs time.

### 7.2. The exponential law model

$$v_s = \frac{dp}{d\rho} = 1 \qquad (60)$$

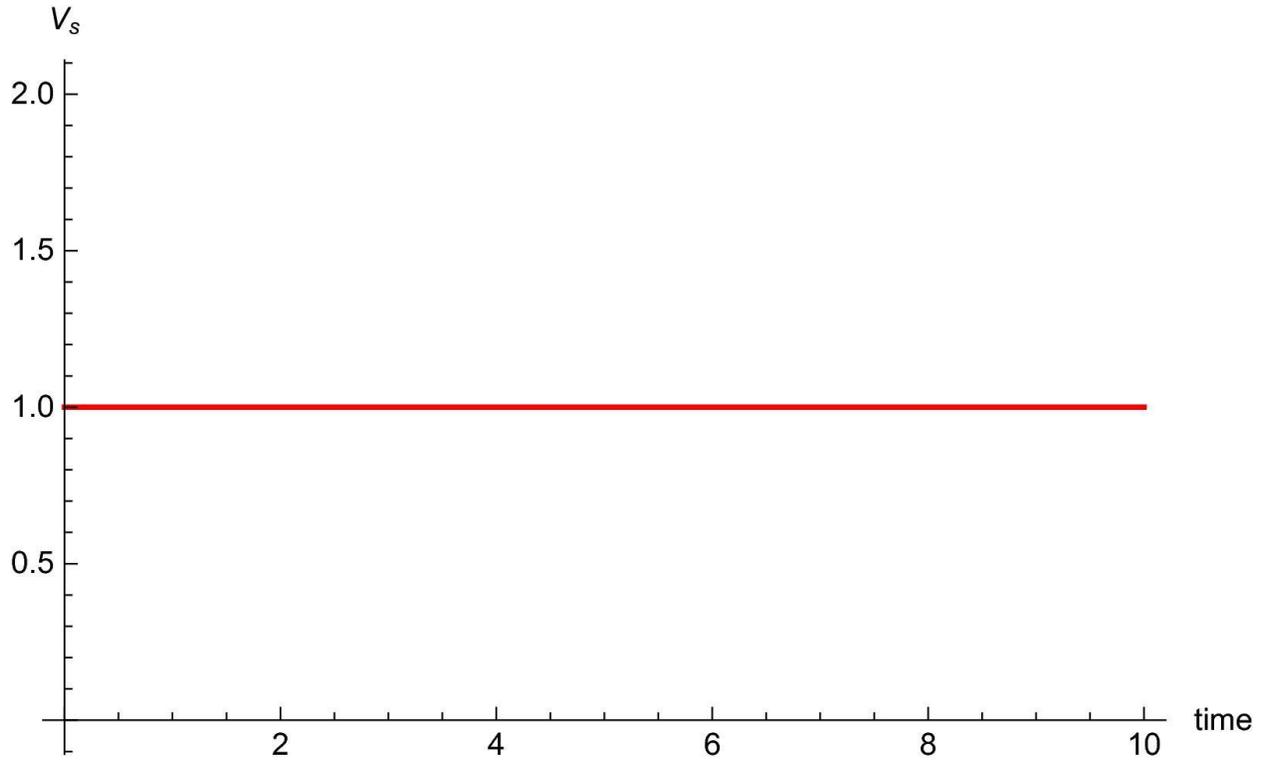

Figure 15. $v_s$ vs time.

It is observed from figure 14 that the function is positive at the early stages of the evolution of the universe and tends to be negative at late time. Thus, the model is stable at the early phases of the universe and is unstable at late times. Therefore after the Big Bang there is instability. Figure 15 shows the behavior of the $v_s$ with time. It is clear that the function is positive and the model is stable.

## 8. The Statefinder diagonsis:

The statefinder pair $\{r,s\}$ is defined as ((Sahni et al. (2003), Akarsu et al.(2014))

$$r = \frac{\dddot{R}}{RH^3} = q + 2q^2 - \frac{\dot{q}}{H}, \quad s = \frac{r-1}{3\left(q - \frac{1}{2}\right)}. \qquad (61)$$

## 8.1. Power law model

The statefinder parameters are found as

$$r = \frac{n^2 - 9n + 18}{n^2} \text{ so that } s \text{ becomes } s = \frac{2}{n}$$

The relationship between the two statefinder parameters is

$$r = \frac{9s^2 - 9s + 2}{2}. \tag{60}$$

For $(r,s) = (1,0)$ indicates that the $\Lambda$CDM model whereas we get the CDM model for $(r,s) = (1,1)$. Hence the derived model corresponds with the value of $\Lambda$CDM limit which resembles with current observations and investigations.

## 8.2. Exponential Law model

The statefinder parameters are found as

$$r = 1 \text{ and } s = 0 \tag{62}$$

The $\Lambda$CDM model corresponds to $r = 1$ and $s = 0$. Hence, after proper analysis, it is observed that there is coincidence with the flat $\Lambda$CDM model in the future by our domain walls model (Shaikh et. al. 2019).

## 9. The Jerk Parameter:

The jerk parameter in cosmology is defined as a dimensionless third derivative of the scale factor with respect to the cosmic time (Chiba and Nakamura (1998), Blandford et al. 2004, Visser (2004, 2005)) as

$$j(t) = \frac{1}{H^3}\frac{\dddot{R}}{R} = q + 2q^2 - \frac{\dot{q}}{H}, \qquad (63)$$

where $R$ is the cosmic scale factor, $H$ is the Hubble parameter and the dot denotes differentiation with respect to the cosmic time.

### 9.1. Power Law Model

The jerk parameter yields

$$j(t) = \frac{n^2 - 9n + 18}{n^2} \qquad (64)$$

Equation (64) gives a positive value for an appropriate choice of $n$. Thus there is a smooth transition of the universe from decelerating to accelerating phase of the universe ( Shaikh et. al. 2020).

### 9.2. Exponential Law Model

The jerk parameter yield

$$j(t) = 1 \qquad (65)$$

This value overlaps with flat Lambda CDM models. For this Universe it turns out to be constant. Sahoo and Sivakumar (2015) presented the models closer to the Lambda CDM model twice in the evolution .

## 10. Some observational constraints

In this section, the consistency of our model with the observational parameters is investigated. The physical parameters such as red shift, look-back time, proper distance, luminosity distance, angular distance etc. are evaluated.

### 10.1. Exponential Law Model:

**Look-back time-red shift:**

The look-back time is defined as the elapsed time between the present age of the Universe and the age of the universe, when a particular light from a cosmic time source at a particular redshift $z$ was emitted. It is defined as

$$t_L = t_0 - t = \int_R^{R_0} \frac{dR}{R}, \tag{66}$$

where $R_0$ is the present day scale factor of the universe and

$$\frac{R_0}{R} = 1 + z. \tag{67}$$

For the discussed model, we have

$$\frac{R_0}{R} = 1 + z = \left(\frac{e^{t_0}}{e^t}\right)^{\frac{1}{3}} \tag{68}$$

The above equation takes the form

$$H_0(t_0 - t) = \log(1+z),$$

$$t = t_0 - \frac{\log(1+z)}{H_0}, \tag{69}$$

where $H_0$ is the present Hubble's parameter.

**Proper distance:**

The proper distance $d(z)$ is defined as the distance between a cosmic source emitting light at any instant $t = t_1$ located at $r = r_1$ with redshift $z$ and the observer receiving the light from the source emitted at $r = 0$ and $t = t_0$. Thus

$$d(z) = r_1 R_0, \text{ where } r_1 = \int_t^{t_0} \frac{dt}{R} = H_0^{-1} R_0^{-1} z. \tag{70}$$

For the discussed model, we have the proper distance as

$$d(z) = H_0^{-1} z. \tag{71}$$

The proper distance $d(z)$ is linear function with respect to the redshift $z$. It is observed that $d(z = \infty)$ is always infinite.

**Luminosity distance:**

The apparent luminosity of a source at radial coordinate $r_1$ with a redshift $z$ of any size $d_L$ is defined as

$$d_L = \left(\frac{L}{4\pi l_*}\right)^{\frac{1}{2}} = r_1 R_0 (1+z), \qquad (72)$$

where $L$ is the absolute luminosity distance and $l_*$ is the apparent luminosity of source. Using equations (70) and (72), it yields

$$d_L = d(Z)(1+z). \qquad (73)$$

For the discussed model, using equations (71) and (73), we have luminosity distance $d_L$ as

$$d_L = H_0^{-1} z(1+z), \qquad (74)$$

which shows that the luminosity distance increases faster with redshift $z$. For early inflationary epoch i.e. $z \to \infty$, this does not remain finite while for late time universe, i.e. $z \to -1$, this tends to zero.

**Angular diameter:**

The angular diameter of a light source of diameter $D$ at $r = r_1$ and $t = t_1$ observed at $r = 0$ and $t = t_0$ is given by

$$\delta = \frac{D}{r_1 R(t_1)} = \frac{D(1+z)^2}{d_L}. \qquad (75)$$

The angular distance $d_A$ is defined as the ratio of the source diameter to its angular diameter

$$d_A = \frac{D}{\delta} = d_L(1+z)^{-2}. \tag{76}$$

Using equation (74) for the presented model, we have

$$d_A = H_0^{-1} z(1+z)^3. \tag{77}$$

### 10.2. Power Law Model:

**Look-back time:**

$$H_0(t_0 - t) = \frac{n}{3}\left[1 - (1+z)^{\frac{-3}{n}}\right]. \tag{78}$$

For n = 2, it describes the well-known look back time in Einstein-de sitter Universe.

For small $z$, we have

$$H_0(t_0 - t) \approx 0. \tag{79}$$

For large $z$, we have

$$H_0(t_0 - t) \approx \frac{n}{3}. \tag{80}$$

**Proper distance:**

$$d(z) = a_1 H_0^{-1}\left[1 - (1+z)^{1-\frac{3}{n}}\right]. \tag{81}$$

The proper distance at an initial epoch $d_{z \to \infty} = a_1 H_0^{-1}$ and $d(z \to -1)$ is always infinite for late universe.

**Luminosity distance:**

$$d_L = a_1 H_0^{-1}\left[1-(1+z)^{1-\frac{3}{n}}\right](1+z). \tag{82}$$

**Angular diameter:**

$$d_A = d_L(1+z)^{-2} = a_1 H_0^{-1}\left[1-(1+z)^{1-\frac{3}{n}}\right](1+z)^{-1}. \tag{83}$$

The angular parameter $d_A$ approaches the finite constant $a_1 H_0^{-1}$ as for the large value of red shift i.e. $z \to \infty$ with respect to the condition $n > 1$. Hence objects with large redshift look very faint.

## 11. Discussion and Concluding remarks

The purpose of this paper is to investigate the recently developed $f(T)$ gravity. For this purpose LRS Bianchi type I space-time model is considered in the presence of domain walls and the solutions of the field equations has been investigated by obtaining two different types of models such as power law and exponential law.

- It is found that, power law model has an initial singularity while exponential model is free from any type of singularity.
- The underlying physical consequence is that the decreasing of matter energy density with increase of cosmic time leads to the expanding universe.
- In power law model, after the Big Bang there is instability while the model is stable in exponential law.
- Jerk parameter and statefinder trajectory in the r- s plane are close to $\Lambda$CDM model.

- It is observed that the luminosity distance increases linearly with red-shift ($z$) for small value of red-shift ($z$).

The derived models throw some light in the understanding of structure formation of the universe in modified theory of gravity.